\documentstyle{article}
\def\bc{\begin{center}}
\def\ec{\end{center}}
\def\be{\begin{eqnarray}}
\def\ee{\end{eqnarray}}

\def\d#1#2{\frac{\displaystyle #1}{\displaystyle #2}}

\def\r{\partial}

\def\al{\alpha}
\def\eps{\epsilon}

\def\ka{\kappa}

\def\om{\omega}

\def\PRA{{\it Phys. Rev.}~{\bf A}}

\def\PLA{{\it Phys. Lett.}~{\bf A}}

\begin{document}
%\hspace {3in}

%\vfill
\bc{\Large \bf Poynting vector, energy density and energy velocity in anomalous dispersion medium}
\ec

\bigskip

\bc {Chao--Guang Huang$^{a,c}$ and Yuan--Zhong Zhang$^{b,c}$
\hspace{2cm}

{\small {$^a$ \it Institute of High Energy Physics, Chinese Academy of Sciences ,
P.O.Box 918, Beijing 100039, China}}

{\small {$^b$ \it CCAST (World Laboratory), P.O. Box 8730, Beijing 100080}}

{\small {$^c$ \it Institute of Theoretical Physics, Chinese Academy of Sciences,
P. O. Box 2735, Beijing 100080, China}}
}\ec

%\vfill
\bigskip\bigskip

\bc
\parbox{13cm}
{{\bf Abstract.}\quad  The Poynting vector, energy density and energy velocity of light pulses
propagating in anomalous dispersion medium (used in WKD-like experiments) are calculated. Results 
show that a negative energy density in the medium propagates along opposite of incident direction 
with such a velocity similar to the negative group velocity while  the direction of the Poynting
vector is positive. In other words, one might say that a positive energy density in the medium 
would propagate along the positive direction with a speed having approximately the absolute value
of the group velocity.
We further point out that neither energy velocity nor group velocity is a good concept to describe 
the propagation process of light pulse inside the medium in WKD experiment owing to the strong 
accumulation and dissipation effects.
}
\ec
\bigskip
%\vspace{2cm}
\bc
\parbox{13cm}
\ec

%\vfill

%\begin{figure}[b]
%\rule[-2.5truemm]{5cm}{0.1truemm}\\[2mm]
%{\footnotesize
%1.}
%\end{figure}

%\pagebreak
\bigskip\bigskip

\quad ~ Wang, Kuzmich and Dogaiu claim that they have observed the superluminal light
propagation by sending the probe laser pulses through the particularly prepared atomic
caesium vapour and measuring the deviation of the refractive index of the medium
from that of the vacuum and the arrival time of the pulses\cite{{WKD},{DKW}}.  Their main
conclusion is that the light pulse propagates through the medium with group velocity about
$-c/310$.  Some comments on the WKD experiment have been made [3--6]
in different ways.  Recently, we analyse the negative group velocity and distortion of
light pulse in the anomalous dispersion medium with two gain lines and show
that although the shape of the light pulses seems to be almost preserved in a specific view,
the distortion is still too large to use the concept of the group velocity to describe the
propagation of the light in the medium\cite{HZ}.  In the present letter, we shall study
the Poynting vector, energy density and energy velocity of light pulse in the anomalous
dispersion medium.

Without loss of generality, let the incident light pulse go to the medium at $z=0$ from vacuum in $z<0$.
The electric and magnetic fields ${\bf E}(r,t)$ and ${\bf H}(r,t)$ of the light
pulse propagating in the medium in the $z$-direction can be expanded by complex Fourier series as
\be
{\bf E}(z,t) = \d 1 {2\sqrt{2\pi}}\left (\int d\om {\bf E}(z,\om) {\rm e}^{-i(\om + \om_c) t}
+{\rm c.c.}\right )
\ee
and
\be
{\bf H}(z,t) = \d 1 {2\sqrt{2\pi}}\left (\int d\om {\bf H}(z,\om) {\rm e}^{-i(\om +\om_c)t}
+{\rm c.c.}\right ),
\label{mf}
\ee
where $\om_c$ is the angular frequency of the carrier wave, $\om$ is the {\it difference} between the
angular frequencies of the Fourier component and the carrier wave, ${\bf E}(z,\om)$ and ${\bf H}(z,\om)$
are the complex Fourier components of the electric and magnetic fields, respectively,
with the angular frequency $\om +\om_c$, and c.c. stands for the complex conjugate.  For monochromatic
plane wave with angular frequency $\om +\om_c$, we have
\be
{\bf H}(z,\om) = \sqrt {\eps (\om)} {\bf e}_k \times {\bf E}(z,\om),
\label{rHE}
\ee
where $\eps (\om)$ is the dielectric permeability, and ${\bf e}_k$ is the unit vector of wave.
We have here chosen the magnetic conductivity $\mu =1$ and then the refractive index is $n(\om)=
\sqrt {\eps (\om)}$.  Note that {\bf E}, {\bf H} and ${\bf e}_k$ are perpendicular to each other and form
a right-handed system.

Substituting the above equations in the Poynting vector
\be
{\bf S} = \d c {4\pi} ({\bf E} \times {\bf H})
\label{Pv}
\ee
and taking the average over a period of carrier wave, we obtain
\be
{\bar {\bf S}} =\d {c {\bf e}_k} {32 \pi ^2} {\rm e}^{-2{\rm Im}(k(0))z} \left (\int d\om A(z,\om)
{\rm e}^{-i\om t} \int d\om' n^*(\om') A^*(z,\om') {\rm e}^{i\om' t} +  {\rm c.c} \right ),
\ee
where $k(0)$ is the wave number of the carrier wave, and $A(z,\om)={\bf e}_E \cdot {\bf E}(z,\om){\rm e}^{-ik(0)z}$.
Note that
\be
A(z,\om)=A(0,\om) \exp [i(k(\omega) -k(0)) z]
\ee
with $k(\om)=(\om +\om_c)n(\om)/c$ and that
\be
k(\om)- k(0) =\ka _1 \om + \d 1 2 \ka _2 \om^2 + ...,
\label{k}
\ee
where $\ka_l=(\r^l k(\om)/\r\om ^l)_0$ with $l=1, 2, ...$.
The subscript 0 stands for taking value at zero-deviation from the carrier frequency.
Thus, in order to obtain the Poynting vector, one has to calculate the integrals
\be
I_1 =\d 1 {\sqrt {2\pi}} \displaystyle \int ^{+\infty }_{-\infty } d\om A(0,\omega )
\exp [-i\omega t +i\ka_1 \omega z +i \ka_2 \om^2 z/2+...]
\label{I1}
\ee
and
\be
I_2 =\d 1 {\sqrt {2\pi}} \displaystyle \int ^{+\infty }_{-\infty } d\om n(\om) A(0,\omega )
\exp [-i\omega t +i\ka_1 \omega z +i \ka_2 \om^2 z/2+...].
\label{I2}
\ee

Now, consider the Guassian pulse, whose envelop at $z=0$ has the form
\be
A(0,t)=A_0 \exp [- \al ^2(t-t_0)^2/2],
\label{gp}
\ee
where $\al$ and $t_0$ are two real constants, which determine the width and maximum position of the pulse
respectively.  Its Fourier transformation is
\be
A(0,\om)=\d 1 \al \exp [-\om ^2/(2 \al^2)].
\label{gpft}
\ee
Keeping $\ka_2$ term and neglecting the higher order terms, one gets
\be
I_1 = \d {A_0} {\sqrt{1-i\ka_2 \al^2 z}} \exp \left [-\d {\al^2}2 \d {(t-t_0-\ka_1z)^2}
{1-i\ka_2 \al^2 z} \right ],
\label{Azg}
\ee
\be
I_2= g I_1 ,
\ee
where
\be
g= n (\om _s) +\d 1 2 \d {n''(0) \al^2} {1-i\ka_2 \al ^2 z},
\ee
in which $n''(0)=(\r^2 n(\om)/\r \om^2)_{\om=0}$ is the second derivative of refractive index with respect to
$\om$ at zero-deviation from the carrier frequency, and
\be
\om_s = - i\d {\al ^2(t-t_0 -\ka_1z)} {1-i \ka_2 \al ^2 z}.
\ee
Thus, the mean Poynting vector is
\be
\bar {\bf S}=\d {c{\bf e}_k} {8 \pi } \d {A_0^2} {|1-i\ka_2 \al^2 z|} {\rm Re}(g) \exp \left [
-2{\rm Im}[k(0)]z-\al^2 {\rm Re} \left (\d {(t-t_0-\ka_1z)^2} {1-i\ka_2 \al^2 z}\right ) \right ].
\label{aP}
\ee

The intensity of light is given by
\be
I= \bar {\bf S} \cdot {\bf e}_k {\cal A},
\ee
where ${\cal A}$ is the area of the cross-section of the light beam.  It is evident that the intensity of light
$I>0$ as long as Re$(g)>0$.  When ${\rm Re}(g) \approx 1$, the intensity of light reduces to
\be
I \propto \d 1 {|1-i\ka_2 \al^2 z|} \exp \left [-\al^2 {\rm Re} \left (
\d {(t-t_0-\ka_1z)^2} {1-i\ka_2 \al^2 z}\right ) \right ] ,
\label{ril}
\ee
which is the same as Eq.(31) in \cite{HZ} except the definition of $\ka_1$ here is equal to the one in \cite{HZ}
plus $1/c$.

In WKD experiment\cite{{WKD},{SPH}}, $\om_c =7\pi \times 10^{8}$MHz, $\al =0.45$MHz and
\be
n(\om)=\left \{ 1 + 4 \pi m \left [\d 1 {\om -\om_0+ i\beta} +
\d 1 {\om+ \om_0+ i\beta} \right ]\right \}^{1/2},
\label{ri}
\ee
where $\om_0 =2.7\pi$MHz is a half of the angular frequency space of the two gain lines, $\beta=0.92\pi$MHz
is the angular frequency width of the gain lines, and $m=0.36\pi$Hz is related to the gain coefficients.
Neglecting
\be
\d {m^2} {\om_0^2 + \beta^2}, ~ \d {m^3 \om_c} {(\om_0^2 + \beta^2)^2}, ~\d m {\om_c}
\label{neg}
\ee
and higher order terms, we obtain\cite{note}
$\ka_1 \approx -1.026 \times 10^{-8} - 5.276 \times 10^{-15}i {~\rm s (cm)^{-1}}$,
$\ka_2 \approx 1.965 \times 10^{-22}-2.415 \times 10^{-15}i  {~\rm s^2 (cm)^{-1}}$, and Re$(g)=1$.
Thus, the direction of the energy current in the medium is the same as the incident one of the  pulse,
and Eq.(\ref{ril}) is a good approximate expression for the intensity of light.

Now, let's turn to study the electromagnetic energy density and energy velocity in the medium.  Using Maxwell's
equations, the time average of divergence of Poynting vector over a period of the carrier wave can be written
as
\be
- \overline {\nabla \cdot {\bf S}} = \d 1 {4\pi} \left ( \overline {{\bf E} \cdot \d {\r \bf D}{\r t}} +
\overline {{\bf H} \cdot \d {\r \bf B}{\r t}}\right ) =  \d 1 {4\pi} \left (\overline {{\bf E} \cdot
\d {\r \bf D}{\r t}} + \d 1 2 \overline {\d {\r ({\bf H} \cdot {\bf H})}{\r t}}\right ).
\label{divS}
\ee
Similar to ${\bf E}(z,t)$ and ${\bf H}(z,t)$, the induction ${\bf D}(z,t)$ can be expanded by complex
Fourier series as
\be
{\bf D}(z,t) = \d 1 {2\sqrt{2\pi}}\left (\int d\om \eps (\om) {\bf E}(z,\om) {\rm e}^{-i(\om + \om_c) t}
+{\rm c.c.}\right ).
\ee
Thus,
\be
\d {\r \bf D}{\r t} =\d 1 2 \left (-i f(0){\bf A}(z,t) + f'(0) \d {\r {\bf A}(z,t)}{\r t}
+\d i 2 f''(0) \d {\r ^2 {\bf A}(z,t)}{\r t^2} + ... \right )e^{-i\om_c t + ik(0)z} + {\rm c.c.} ,
\ee
where $f(\om)=(\om+\om_c)\eps(\om)$, the prime stands for the derivative with respect to $\om$, and
${\bf A}(z,t)={\bf E}(z,t)\exp(i\om_c t - ik(0)z)$ is the complex envelop of electric field strength
${\bf E}(z,t)$.  The time average of the first term on the right-hand side of Eq.(\ref{divS}) in a
period of carrier wave is
\be
\overline {{\bf E} \cdot \d {\r \bf D}{\r t}} =\d 1 4 \d {\r [{\rm Re}[f'(0)]{\bf A}(z,t)
\cdot {\bf A}^*(z,t){\rm e}^{-2{\rm Im}[k(0)]z}]}{\r t} + 4 \pi Q ,
\label{1stt}
\ee
where
\be
Q = \d 1 {16\pi} \left \{ 2\om_c {\rm Im}(\eps (0)) |{\bf A}(z,t)|^2 + 2 {\rm Im}[f'(0)]
{\rm Im}\left ({\bf A}(z,t) \cdot \d {\r {\bf A}^*(z,t)}{\r t}\right) + \right .
\nonumber \\
\left . {\rm Im} \left [{\bf A}(z,t) \cdot \left ( f''(0) \d {\r ^2 {\bf A}(z,t)}{\r t^2} \right )^*
\right ] + ... \right \} {\rm e}^{-2{\rm Im}[k(0)]z}.
\label{Ga}
\ee
Substituting Eq.(\ref{1stt}) with (\ref{Ga}) in (\ref{divS}), we have
\be
\d {\r \bar w}{\r t} + \nabla \cdot {\bar{\bf S}} = - Q ,
\label{ec}
\ee
with
\be
\bar w \equiv \d 1 {8 \pi}\left[\d 1 2 {\rm Re}[f'(0)]{\bf A}(z,t) \cdot {\bf A}^*(z,t) {\rm e}^{-2{\rm Im}[(k(0)]z}
+ \overline {{\bf H}(z,t) \cdot {\bf H}(z,t)})\right].
\label{ed}
\ee
It should be noted that the bars denoting the time average can be and hence have been moved inside the derivative 
operators on
the left-hand side of Eq.(\ref{ec}).  It should also be noted that in Eq.(\ref{ed}), ${\bf A}(z,t)$ is the
{\it complex} envelop of the electric field, while ${\bf H}(z,t)$ is the {\it real} strength of magnetic
field given by Eq.(\ref{mf}).  Eq.(\ref{ec}) is the differential conservation equation of the electromagnetic
energy.  $\bar w$ is the difference between the mean internal energy per unit volume with and without the
electromagnetic field\cite{LL}.  It may be interpreted as the electromagnetic energy density in the medium
when the light pulse gets through.  $Q$ is the mean heat evolved per unit time and volume, representing the
absorption (or dissipation) rate of energy per unit volume.  The negative absorption (or dissipation) means that
there exits gain (or accumulation) in the medium.  One can define the velocity of transfer of energy density
({\it i.e.} energy velocity) as
\be
v_e \equiv \d {\bar {\bf S} \cdot {\bf e}_k} {\bar w}.
\ee

For the Guassian pulse Eq.(\ref{gp}), as an example,
\be
\bar w = \d {A_0^2} {16 \pi} \d {{\rm Re}[f'(0)]+|g|^2 } {|1-i\ka_2 \al^2 z|}
\exp \left [-2{\rm Im}(k(0))z-\al^2 {\rm Re} \left (\d {(t-t_0-\ka_1z)^2} {1-i\ka_2 \al^2 z}\right ) \right ].
\label{aed}
\ee
It can be checked that Eq.(\ref{ec}) holds for the Guassian pulse with Eqs.(\ref{Azg}), (\ref{aP}) and (\ref{aed})
if the terms in Eq.(\ref{neg}), the terms containing $f'''(0)$ and higher order terms are neglected.  In the
approximation,
\be
v_e \approx \left (\d {d[(\om+\om_c) {\rm Re}(n)/c]}{d\om}\right )^{-1} = v_g .
\ee
Namely, the energy density transfers with the velocity approximately equal to group velocity.

For WKD experiment,
\be
{\rm Re}[f'(0)]+|g|^2 \approx {\rm Re}(n) {\rm Re}\left (\d {d[(\om+\om_c) n]} {d\om} \right ) \approx -310 < 0,
\ee
so that $\bar w < 0$.  If one interpreted $\bar w$ as the electromagnetic energy density of the pulse in
the medium, then the electromagnetic energy density in the medium would be negative!  Besides, the direction of
increase of energy density in the medium would be the opposite of that of $|{\bf A}(z,t)|$!   The negative
energy velocity $v_e \approx  v_g < 0$ would then imply that the negative energy density propagates in the
opposite of the incident direction.

To recover the positivity of energy density, one can replace $\bar w$ in Eq.(\ref{ec}) by
\be
\bar U = U_0 + \bar w \qquad {\rm with} \qquad U_0 =const. > 0.
\ee
$U_0$ is the zero-point energy density coming from the preparation of the medium so that the atomic
population inverted, the part of electromagnetic energy of two beams for preparing the system being stored
in caesium atoms.  The probe pulse through the medium will cause that the number of inverted atomic
population, so as the total energy density, decreases at some time.  This is in accordance with the fact
that the negative oscillator strength is assigned to obtain Eq.(\ref{ri}) \cite{{DKW},{Chiao}}.  $U_0$
also sets the upper limit of the intensity of light.  It reflects the possible saturation
effect\cite{{WKD},{DKW}}.  The negative energy density propagating in the opposite of the incident direction 
is equivalent to the positive energy density propagating with $|v_e|<c$ along the incident direction at
any $z$ in the medium.  Mathematically, one may define the energy velocity in this case as
\be
v_e \equiv \d {\bar {\bf S} \cdot {\bf e}_k} {U_0 - \bar U} = - \d {\bar {\bf S} \cdot {\bf e}_k} {\bar w}
\approx -v_g >0.
\ee
It implies that the propagation of positive energy density is subluminal at every point.

It seems that there is a contradiction between the subluminal propagation of positive energy density
and the peak of pulse arriving at the exit of the medium earlier than through the vacuum, even earlier
than arriving at the entrance of the medium.  However, the contradiction would exist only if $Q$ could be
negligible.  In fact, $Q$ in WKD experiment cannot be neglected. For the Gaussian pulse
Eq.(\ref{gp}),
\be
\d {\r \bar w}{\r t} = 2 {\rm Im}(\om_s) \bar w.
\ee
${\rm Im}(\om_s)$ depends on $t$ and $z$.  Its typical value for the pulse is about $10^6 {\rm s^{-1}}$.  While
\be
Q \approx  \left \{2\om_c {\rm Im}(\eps (0)) + {\rm Im} [ f''(0)] {\rm Re} \left (\d {\al^2} {1-i\ka_2 \al^2 z}
\right ) + {\rm Im} [ f''(0)] {\rm Re} \left (\om_s^2 \right ) \right \} \d {\bar w}
{{\rm Re}(f'(0))+|g|^2 },
\label{Ga2}
\ee
where the second term in Eq.(\ref{Ga}) has been neglected because it is much smaller than the other two terms.
The first term in the braces in Eq.(\ref{Ga2}) gives a constant gain (or accumulation) rate, which is three
orders of magnitude greater than the typical value of $\om_s$.  It magnifies the intensity of light by
$\exp\{-2{\rm Im}[k(0)]z\}$ times.  The second term, depending on $z$, and the third term, depending on
both $t$ and $z$, contribute the gain (or accumulation) and absorption (or dissipation) rates, respectively,
and are about two orders of magnitude greater than the typical value of $\om_s$.  It is the latter two terms that
are responsible for the advancement shift of the pulse.  The second term shows that the gain rate increases
as $z$ increases.  The third term shows that the absorption rate takes the minimum value at $t=t_0 + \ka_1z$.
Although the absorption rate is time-symmetric around $t=t_0 + \ka_1z$, the effect is very different because
the intensity of light is very small when $t<t_0+\ka_1z$ and becomes larger when $t>t_0+\ka_1z$.  In words,
the gain (or accumulation) rate in WKD-like experiments are not constant as expected in \cite{DKW}.

Combining the study on the propagation of the wave front with a step-function in \cite{HZ}, we can now give
the following picture for the WKD experiment:  The leading edge of the Guassian pulse entering the medium
induces the gain (or accumulation) effect that makes the peak of the pulse to appear at the exit of
the medium earlier than it through the vacuum.  At later time when the peak of the pulse reaches
the medium, the absorption (or dissipation) effect becomes more important, and so that a part of the pulse
energy would be dissipated inside the medium.  Thus the peak at the exit of the medium should not be the
entrance peak.  This picture is in agreement with Zhang's argument about gain effect\cite{Zhang2}, McDonald's
energy loan argument\cite{McD}, as well as Sprangle-Pe\~nano-Hafizi's differential gain argument\cite{SPH}.
In other words, the process of the pulse through the caesium vapour is not a simple propagation process of
energy because there exists strong interaction between the light pulse and the medium.  In this sense,
the energy velocity as well as the group velocity are all not good concepts of describing the process of
light pulse propagating through the medium in WKD experiment.

\bigskip

\section*{Acknowledgements}
One of the authors, Y.Z. Zhang, would like to thank Dr. K.T. McDonald for helpful discussion on the negative group velocity and 
energy velocity.
This project is in part supported by the Ministry of Science and Technology of China under grants 95-Yu-08,
NKBRSF G19990754, and National Natural Science Foundation of China under Grant Nos. 19675046, 10047004, 19835040.

\end{document}